\documentclass[runningheads]{llncs}
\usepackage[T1]{fontenc}

\usepackage{booktabs}
\usepackage{tabularx}
\usepackage[separate-uncertainty=true,table-figures-decimal={2}]{siunitx}

\usepackage{graphicx}
\usepackage{subcaption}
\usepackage{hyperref} 
\usepackage[htt]{hyphenat}
\usepackage{kotex}
\usepackage{amsmath}
\usepackage{color}
\usepackage{multirow}

\newcommand{\ourtool}{\textsc{HT-Verbs}}

\begin{document}
\sloppy
\title{Noisy Neighbor: Exploiting RDMA for Resource Exhaustion Attacks in Containerized Clouds}
\titlerunning{Exploiting RDMA for Resource Exhaustion Attacks in Containerized Clouds}
%
\author{Gunwoo Kim\inst{1} \and
Taejune Park\inst{2} \and
Jinwoo Kim\inst{1}\thanks{Corresponding author}
}
\authorrunning{G. Kim et al.}
%
\institute{Kwangwoon University, Seoul, Republic of Korea \\
\email{kgwo0528@gmail.com, jinwookim@kw.ac.kr}\\
Chonnam National University, Gwangju, Republic of Korea \\ \email{taejune.park@jnu.ac.kr}
}

\maketitle              


%
\begin{abstract}
In modern containerized cloud environments, the adoption of RDMA (Remote Direct Memory Access) has expanded to reduce CPU overhead and enable high-performance data exchange. Achieving this requires strong performance isolation to ensure that one container’s RDMA workload does not degrade the performance of others, thereby maintaining critical security assurances. However, existing isolation techniques are difficult to apply effectively due to the complexity of microarchitectural resource management within RDMA NICs (RNICs). This paper experimentally analyzes two types of resource exhaustion attacks on NVIDIA BlueField-3: (i) state saturation attacks and (ii) pipeline saturation attacks. Our results show that state saturation attacks can cause up to a 93.9\% loss in bandwidth, a 1,117× increase in latency, and a 115\% rise in cache misses for victim containers, while pipeline saturation attacks lead to severe link-level congestion and significant amplification, where small verb requests result in disproportionately high resource consumption. To mitigate these threats and restore predictable security assurances, we propose \ourtool{}, a threshold-driven framework based on real-time per-container RDMA verb telemetry and adaptive resource classification that partitions RNIC resources into hot, warm, and cold tiers and throttles abusive workloads without requiring hardware modifications.

\end{abstract}

\section{Introduction}
RDMA (Remote Direct Memory Access) is a high-performance technology that enables direct memory access between systems without involving the host CPU or operating system kernel, significantly reducing latency and CPU overhead while delivering high throughput~\cite{guo2016rdma}. RDMA communicates through RDMA NICs (RNICs), such as the NVIDIA BlueField-3, which are designed with programmable cores integrated into the host NIC system on chip. These cores are directly connected to data center Ethernet or InfiniBand links, allowing RNICs to process packets on-path and respond without involving the host OS networking stack, thus reducing latency for certain applications~\cite{haecki2022diagnose,le2017uno}. Based on these advantages, RNICs are widely adopted in recent cloud environments, including distributed machine learning frameworks and high performance storage systems, where low-latency and high throughput communication are critical~\cite{xiao2024conspirator,liu2019offloading}.

Despite these advantages, RNICs face a significant challenge in cloud environments: the \emph{performance isolation} problem. This issue arises when a malicious tenant exhausts RNIC microarchitecture resources (e.g., NIC caches, processing units) by generating abusive RDMA workloads. Specifically, an attacker can overload the RNIC through carefully crafted RDMA operations, resulting in increased latency, elevated cache miss rates, and, in severe cases, denial of service (DoS) due to internal resource starvation. Such resource exhaustion disrupts critical NIC components and degrades the performance of co-located tenants by creating system-wide bottlenecks, \emph{thus undermining the security assurance guarantees that multi-tenant clouds depend on to ensure availability and predictable performance}.

The performance isolation issue in RDMA has been explored in several prior studies~\cite{kong2023understanding,lou2024harmonic,298513,khalilov2024osmosis,grant2020smartnic}. However, most of these works focus on virtual machine or bare-metal environments, limiting their applicability to containerized architectures. With the shift in cloud infrastructure toward cloud-native architectures based on containers, there is growing interest in adapting RDMA to these environments~\cite{kim2019freeflow,li2024dockrdma,305973,sun2023tsor}. For example, NVIDIA, a leading RNIC vendor, provides an RDMA plugin for Kubernetes~\cite{kube_sriov}, enabling seamless integration of RDMA capabilities within containers. Furthermore, emerging applications such as distributed deep learning frameworks and data-intensive analytics pipelines increasingly leverage RDMA to achieve low-latency, high-throughput communication in containerized setups~\cite{le2017uno,haecki2022diagnose}. Despite these advancements, the impact of resource exhaustion in container environments remains largely unexplored.

In this paper, we systematically investigate the impact of resource exhaustion attacks on the performance of legitimate containers in a containerized cloud environment. We focus on how a malicious container can drain microarchitectural resources of a shared RNIC in RDMA-enabled settings, significantly degrading the performance of co-resident victim containers. To this end, we propose two classes of attacks: (i) \emph{state saturation attacks}, in which the attacker floods RNIC resources such as queue pairs and caches; and (ii) \emph{pipeline saturation attacks}, in which the attacker abuses RDMA verbs to overload the communication pipeline or induce amplified load on the RNIC.

We conduct experiments on NVIDIA BlueField-3, a widely used RNIC, within RDMA-enabled container environments built using Docker and SR-IOV. Our results demonstrate that excessive RDMA operations from an attacker container can cause severe performance degradation in co-located containers. In state saturation attacks, the victim container experiences a bandwidth drop of approximately 93.9\%, a latency increase of up to 1,117×, and a 115\% rise in cache miss rates. Pipeline saturation attacks generate over 20,000 PAUSE frames, indicating link-level congestion, and result in substantial amplification—e.g., 23.1 bytes received by the RNIC per 8-byte request—showing that the attacker consumes disproportionately more resources than the traffic it injects. To mitigate these issues, we propose \ourtool{}, a defense system that enforces fair resource usage and ensures predictable performance in containerized RDMA-enabled environments.\\

\noindent\textbf{Contributions.} We summarize our main contributions as follows:
\begin{itemize}
    \item We present novel resource exhaustion attacks targeting RDMA-enabled container environments, demonstrating how microarchitectural resource contention within RNICs can violate performance isolation guarantees and undermine critical security assurances in multi-tenant clouds.
    \item Through extensive experiments on NVIDIA BlueField-3, we show that our attacks can severely degrade victim container performance, resulting in up to a 93.9\% loss in bandwidth, a 1,117× increase in latency, and a 115\% rise in cache miss rates, thereby quantitatively exposing vulnerabilities that threaten runtime assurance.
    \item We propose \ourtool{}, a conceptual RNIC-based defense mechanism that mitigates these attacks by dynamically monitoring and classifying per-container RDMA resource usage. While a full implementation is left to future work, our design demonstrates the feasibility of enforcing hardware-level resource isolation to restore predictable performance and strengthen security assurances in multi-tenant container environments.
\end{itemize}

\begin{figure}[t]
    \centering
    \includegraphics[width=0.5\linewidth]{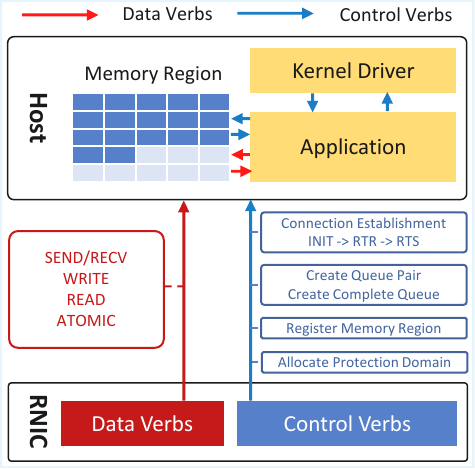}
    \caption{Overview of RDMA workflow.}
    \label{fig:Fig1}
\end{figure}

\section{Background and Motivation}

\subsection{Remote Direct Memory Access (RDMA)}
\label{subsec:rdma}

Fig. \ref{fig:Fig1} illustrates the overall RDMA workflow.
RDMA communication is facilitated through an API called \emph{verbs}, which are categorized into \emph{control verbs} and \emph{data verbs}. Control verbs manage the initialization process by creating and configuring key resources, such as queue pairs (QPs) and completion queues (CQs). Subsequently, the application allocates memory in the host’s DRAM, maps virtual addresses to physical ones, and registers these memory regions with the RNIC (RDMA NIC). This registration enables the RNIC to directly access memory without CPU intervention, thereby reducing latency and improving throughput.

Data verbs, such as WRITE and SEND, are used to actually move bytes between endpoints. Once initialization is complete, an application can leverage data verbs to transfer data between local and remote memory. Data verbs fall into two categories: (i) two‑sided verbs and (ii) one‑sided verbs. In the former, both the sender and the receiver must post work requests. For example, a SEND operation places the payload into a receive buffer that the remote CPU has already advertised. In the latter, after capability negotiation, the initiator can read from, write to, or perform atomic operations on the peer’s memory without further involvement of the remote CPU.

Beyond individual verbs, the transport type itself influences scalability. RDMA defines two transport modes: (i) Reliable Connected (RC) and (ii) Unreliable Connected (UC). RC provides reliable, one‑to‑one communication, similar to TCP, between exactly one local QP and one remote QP. It supports all one‑sided verbs and offers full end‑to‑end reliability, but each additional peer requires an extra QP. In large clusters, the RNIC’s QP capacity becomes a limiting factor. In contrast, UC allows a single QP to communicate with multiple peers, similar to UDP, greatly reducing the number of QPs a thread must maintain and thereby improving scalability. The trade‑off is that UC omits support for one‑sided operations and end‑to‑end reliability.

\begin{figure}[t]
    \centering
    \includegraphics[width=\linewidth]{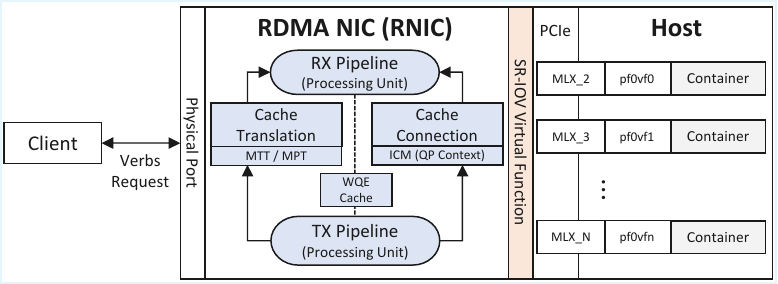}
    \caption{Architecture of an RNIC and an example of SR-IOV-based virtualization to serve multiple containers.}
    \label{fig:Fig2}
    \vspace{-0.2in}
\end{figure}

\subsection{RDMA-enabled Container Environment}

In modern cloud-native infrastructures, containers frequently require direct access to high-performance networking capabilities such as RDMA~\cite{kim2019freeflow}. To support this, a single physical RNIC can be virtualized through SR-IOV (Single Root I/O Virtualization), which provides one Physical Function (PF) for global management and multiple Virtual Functions (VFs), each maintaining its own hardware context (e.g., queue pair and memory translation state). Fig.~\ref{fig:Fig2} shows an example of an RNIC deployment in containerized environments. When VFs (e.g.,~\texttt{pf0vf*}) are created, they are automatically exposed to the host system as distinct PCIe devices (e.g.,~\texttt{mlx5\_*}), reflecting their instantiation as independent virtual ports within the RNIC. By assigning a VF to a container’s network namespace, containers can achieve near-native, low-latency access to the virtualized RNIC.

A physical RNIC consists of various microarchitecture resources~\cite{kong2023understanding}, each dedicated to specific metadata storage and fast access, as shown in Fig.~\ref{fig:Fig2}. For instance, in the case of the NVIDIA BlueField-3, the Memory Translation Table (MTT) handles virtual-to-physical address translation, while the Memory Protection Table (MPT) enforces memory access permissions. The Interconnect Connect Memory (ICM) cache stores QP state and connection metadata to manage RDMA operations. Additionally, the WQE (Work Queue Entry) cache holds pre-fetched entries from transmission and reception queues, enabling rapid processing in the TX/RX pipeline. Transmission requests are routed to the user side through the TX pipeline, while reception requests are delivered to the container’s memory via the RX pipeline.

\begin{figure*}[!t]
    \centering
    \includegraphics[width=0.55\linewidth]{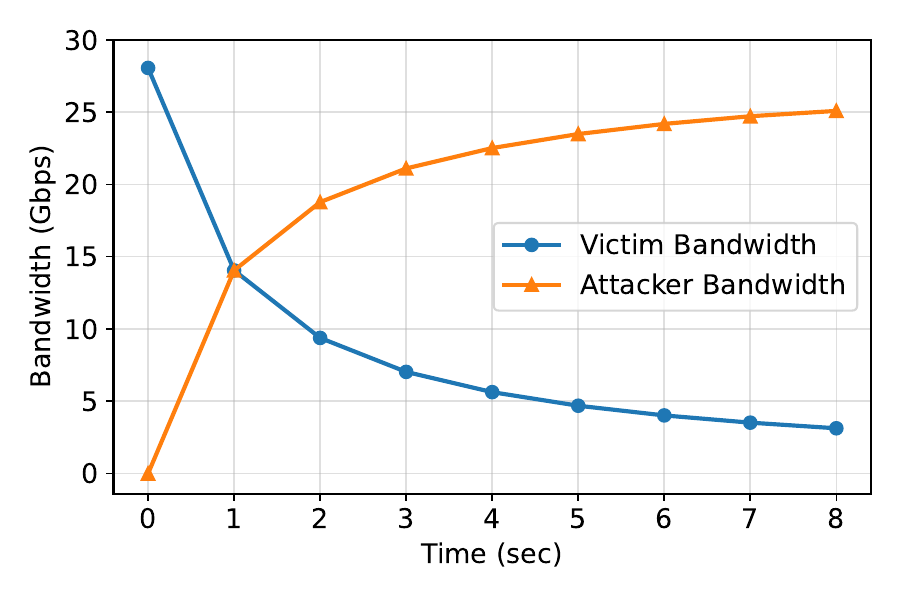}
    \caption{Example of a performance isolation issue where an attacker exhausts the victim's bandwidth by increasing the number of its own QPs.}
    \label{fig:Fig3}
    \vspace{-0.2in}
\end{figure*}

\subsection{Motivating Example}

\emph{Performance isolation} refers to the ability to ensure that one tenant’s activities do not degrade the performance of others sharing the same physical resources~\cite{kong2023understanding}. As discussed earlier, multiple containers in a containerized cloud share a single RNIC’s internal resources. We hypothesize that this setup may inherently suffer from performance isolation issues. To validate this, we conducted an experiment in which two containers—a victim and an attacker—ran concurrently on the same host, each assigned a separate VF on an NVIDIA BlueField‑3. The victim container continuously executed the \texttt{ib\_write\_bw} benchmark to establish baseline throughput, while the attacker container gradually increased its number of QPs from 1 to 8, issuing high-rate WRITE verbs. At each step, we measured the victim’s sustained bandwidth.

As illustrated in Fig.~\ref{fig:Fig3}, increasing the attacker's number of QPs leads to resource saturation within the RNIC—specifically in the TX/RX pipelines and the WQE cache—causing the victim's bandwidth to drop by up to 93.9\%. Interestingly, the total bandwidth was observed to be inversely proportional to the number of QPs created by the attacker, resembling the behavior of fair-share scheduling. Since there is no control over NIC resources in the current RDMA-enabled container environment, these findings reveal that virtualized RNICs in containerized environments are highly susceptible to performance isolation breakdowns, motivating a deeper investigation into RDMA-level attack vectors targeting this vulnerability.
\section{Resource Exhaustion in RDMA-enabled Containers}

\begin{figure}[t]
    \centering
    \includegraphics[width=0.6\linewidth]{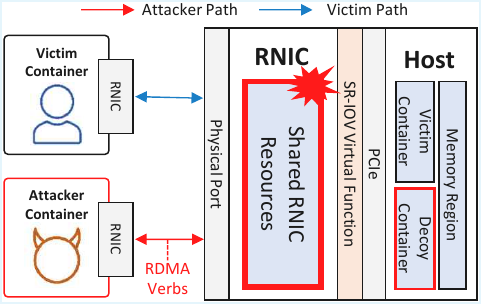}
    \caption{Our threat model, where an \emph{attacker container} compromises the performance isolation of a target host by leveraging a co-resident \emph{decoy container} to launch RDMA-based resource exhaustion attacks against \emph{victim containers}.}
    \label{fig:threat_model}
    \vspace{-0.2in}
\end{figure}

\subsection{Threat Model}
We consider a multi-tenant environment in an on-premises cloud where tenants share the same physical machine. To provide tenants with high-performance applications, cloud administrators may deploy RDMA-enabled containers. This setup is feasible given that Kubernetes supports the SR-IOV plugin~\cite{kube_sriov}, which allows containers to be assigned their own VFs to access RNICs. Accordingly, we assume that the cloud administrator adopts this type of Kubernetes deployment. In this context, we consider an \emph{attacker container} initially residing on a different physical machine from the target host. Fig.~\ref{fig:threat_model} shows our threat model.

The attacker's goal is to induce performance isolation issues on the target host via malicious RDMA operations, thereby disrupting the operations of one or more \emph{victim containers}. To achieve this, the attacker requires a \emph{decoy container} co-residing on the target host. Such co-residency can be achieved through brute-force attacks by repeatedly creating containers combined with co-residency checks, as demonstrated in prior VM-based attacks~\cite{ristenpart2009hey}, and more recently confirmed feasible in container environments~\cite{10.1145/3411495.3421357}. Both the attacker container and the decoy container are assumed to have access to their own VFs assigned via SR-IOV. We further assume that they share the same L3 domain, enabling communication over RoCEv2 (RDMA over Converged Ethernet version 2). Thus, the attacker container can establish a QP with the decoy container, invoke RDMA verbs, and freely choose the transport mode, such as RC or UC (Section~\ref{subsec:rdma}). Note that we do not consider RDMA protocol-related vulnerabilities~\cite{rothenberger2021redmark}, and thus assume that the attacker fully complies with the RDMA protocol and standard operations.

\begin{figure}[t]
    \centering
    \includegraphics[width=1\linewidth,trim={0.2cm 0 0.2cm 0},clip]{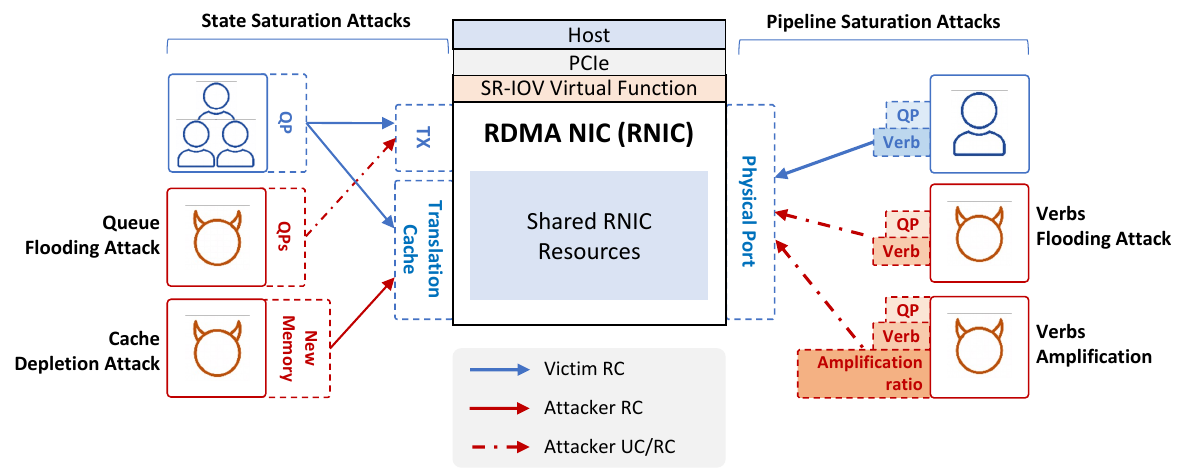}
    \caption{Overview of resource exhaustion attacks in an RDMA-enabled container environment.}
    \label{fig:attack_overview}
    \vspace{-0.2in}
\end{figure}

\begin{table}[b]
    \centering
    \caption{Summary of RDMA resource exhaustion attacks, their target, verbs, and modes.}
    \footnotesize
    \resizebox{\textwidth}{!}{
    \begin{tabular}{c c c c c}
         \toprule
         \textbf{Type} & \textbf{Attack} & \textbf{Target} & \textbf{Verbs} & \textbf{Modes} \\ \midrule

        \multirow{2}{*}{State Saturation Attacks} & Queue Flooding & QPs & SEND, RECV & UC, RC \\ \cmidrule{2-5}
         
        & Cache Depletion & Translation/Connection Cache & READ, WRITE & RC \\ \midrule 
          
        \multirow{2}{*}{Pipeline Saturation Attacks} & Verbs Flooding & WQE, TX/RX pipeline & ALL & UC, RC \\ \cmidrule{2-5}
          
        & Verbs Amplification & QP, TX/RX pipeline  & ALL & UC, RC \\
        \bottomrule
        
    \end{tabular}
    }
    \label{tab:attack_summary}
\end{table}

\subsection{Attack Scenarios}

Our goal is to develop practical attacks that an \emph{attacker container}, in coordination with a \emph{decoy container}, can launch to disrupt performance isolation on the target host by exploiting RDMA operations. To this end, we introduce two types of attacks: (i) \emph{state saturation attacks} and (ii) \emph{pipeline saturation attacks}. Fig.~\ref{fig:attack_overview} provides an overview of these attacks, and Table~\ref{tab:attack_summary} summarizes their targets, RDMA verbs, and operational modes.

\subsection{State Saturation Attacks}

In these types of attacks, the attacker aims to exhaust RNIC resources by aggressively consuming connection management and caching resources involved in RDMA communication.

\vspace{0.05in}

\noindent\textbf{Queue Flooding.} The attacker can rapidly instantiate a large number of queue pairs (QPs) and corresponding completion queues (CQs), similar to a TCP SYN flooding attack. Each new QP allocates on‑chip resources (QP state, work queue buffers), and the flood of posted work requests saturates the TX/RX pipelines, evicts legitimate QP state from the connection cache, and exhausts WQE cache entries. This attack can be launched in both UC and RC modes. In UC mode, the attacker can cycle large batches of SEND/RECV requests on a single QP without requiring collaboration from the decoy container. In contrast, RC mode requires a dedicated pair of QPs between the attacker and decoy containers, making the creation of many QPs more critical—despite the need for decoy container collaboration.

\vspace{0.05in}

\noindent\textbf{Cache Depletion.} Next, the attacker can issue one‑sided READ or WRITE operations with continuously increasing remote addresses across its QPs, flushing the translation cache and connection cache, and driving up cache miss rates on the shared RNIC. The resulting high miss rates introduce delays in the TX/RX pipelines and degrade the performance of co‑located victim containers. If a victim container and a decoy container share cache resources on the RNIC, this attack can severely impact system throughput and significantly increase latency for time-sensitive RDMA workloads on the victim container. Notably, the attacker can achieve similar cache-thrashing effects using a single QP in UC mode by flooding new address operations.

\subsection{Pipeline Saturation Attacks}

In this type of attack, the attacker overwhelms the RDMA processing pipeline by flooding RDMA operations, leading to resource contention and performance degradation.

\vspace{0.05in}

\noindent\textbf{Verbs Flooding.} The attacker can flood RDMA verbs in a manner similar to HTTP or UDP flooding attacks. To do so, the attacker injects verbs at the maximum throughput for each QP and amplifies the impact by increasing the number of QPs filled with verbs. Specifically, in RC mode, the attacker may leverage both one-sided and two-sided verbs across multiple QPs. Each verb invocation allocates per-QP context and triggers reliability handshakes, stressing the atomic engine and TX/RX pipelines, evicting WQE entries and translation cache state, and causing pipeline stalls and context-switch thrashing. In contrast, in UC mode, the attacker uses a single QP to send large bursts of WRITE/SEND messages without acknowledgments; the absence of reliability overhead enables higher injection rates. Consequently, stress on the WQE FIFO and TX/RX pipelines increases rapidly, leading to PCIe back-pressure.

\vspace{0.05in}

\noindent\textbf{Verbs Amplification.} The attacker can design a more sophisticated amplification attack, in which a small RDMA verb request triggers significant resource consumption on the RNIC, similar to traditional DDoS amplification attacks~\cite{kuhrer2014exit}. Specifically, amplification occurs if the RNIC processes more bytes than the payload size of the issued verb. We observe that such amplification arises from additional overhead introduced by protocol headers, control traffic, and flow-control events. Thus, the attacker can exploit this asymmetry to efficiently overload the RNIC while consuming fewer resources.

\section{Evaluation}

In this section, we evaluate the effectiveness of our proposed attacks.

\subsection{Experimental Environment}

To assess the impact of the proposed resource exhaustion attacks, we set up an experimental testbed comprising two servers, each configured with an Intel Core i7-13700K CPU (16 cores @ 3.40 GHz), 64 GB of RAM, and an NVIDIA BlueField-3 RNIC~\cite{bluefield3}. We then replicated an RDMA-enabled container environment with Docker containers, each configured to access one of these VFs (Virtual Functions), ensuring direct, near-native access to the RNIC’s resources. Specifically, on each RNIC, SR-IOV was enabled to instantiate two VFs via \texttt{echo 2 > /sys/class/net/p0/device/sriov\_numvfs}, and the \texttt{pipework} utility~\cite{petazzoni2014pipework} was employed to bind each VF into a separate Docker container network namespace. Attacker containers were designed to launch the proposed resource exhaustion attacks, while victim containers performed normal RDMA operations and were monitored to evaluate the impact of the attacks on their performance. RDMA communication between containers was established over RoCE (RDMA over Converged Ethernet) v2. The attacker containers were programmed to generate excessive RDMA operations to overload the RNIC, while the victim containers followed typical RDMA workload patterns to simulate real-world application behavior.

We measured bandwidth and latency using two benchmark tools: \texttt{ib\_write\_bw} and \texttt{ib\_read\_lat} from the \texttt{perftest} package~\cite{linuxrdmaperftest}, while Linux \texttt{perf} was used to record cache-miss ratios. Additionally, BlueField-3 hardware counters along with its PCIe observability module, were used to quantify per-verb pipeline pressure, enabling to correlate internal RNIC resource contention with the observed bandwidth and latency degradation. Link-level statistics were also collected from the RNIC using the Linux \texttt{ethtool -S} command, monitoring hardware counters, such as \texttt{port\_rcv\_data} and PAUSE frames.

We began by measuring baseline performance—bandwidth, latency, hardware counters, and cache statistics—in the absence of any attacks to establish normal operating metrics for the victim containers. Next, we sequentially executed the queue flooding, cache depletion, and verbs flooding scenarios from the adversary containers, continuously recording the victim’s performance throughout each test. To ensure fair comparison and reproducibility, we kept key variables—such as the total number of containers, RDMA operation types, and data sizes—constant across all experimental runs. This approach minimized external factors influencing the results and ensured reproducibility.

\begin{figure}[t]
    \centering
    \begin{minipage}[t]{0.48\linewidth}
        \centering
        \includegraphics[width=\linewidth]{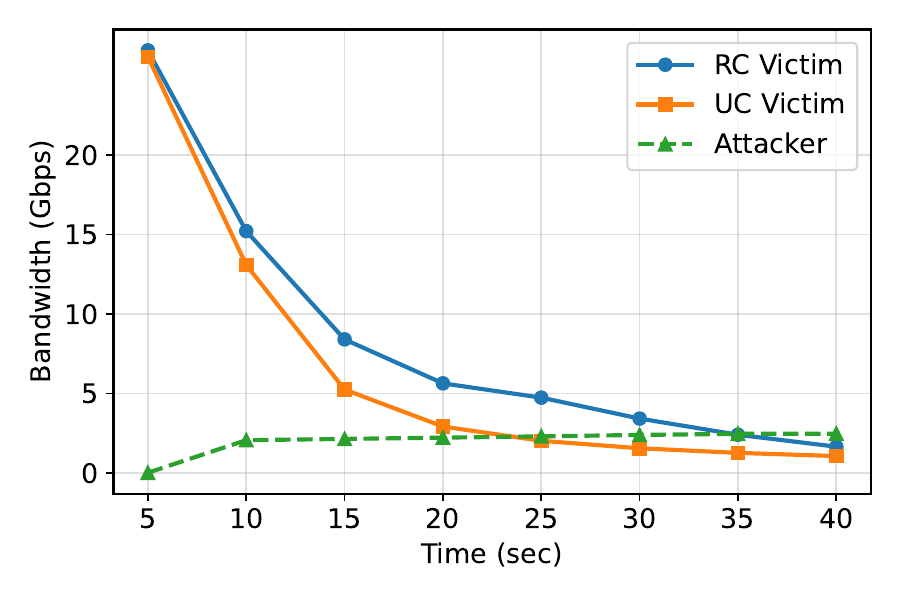}
        \caption{Bandwidth changes in a \emph{single-victim} environment during a queue flooding attack.}
        \label{fig:queue_flood_result}
    \end{minipage}
    \hfill
    \begin{minipage}[t]{0.48\linewidth}
        \centering
        \includegraphics[width=\linewidth]{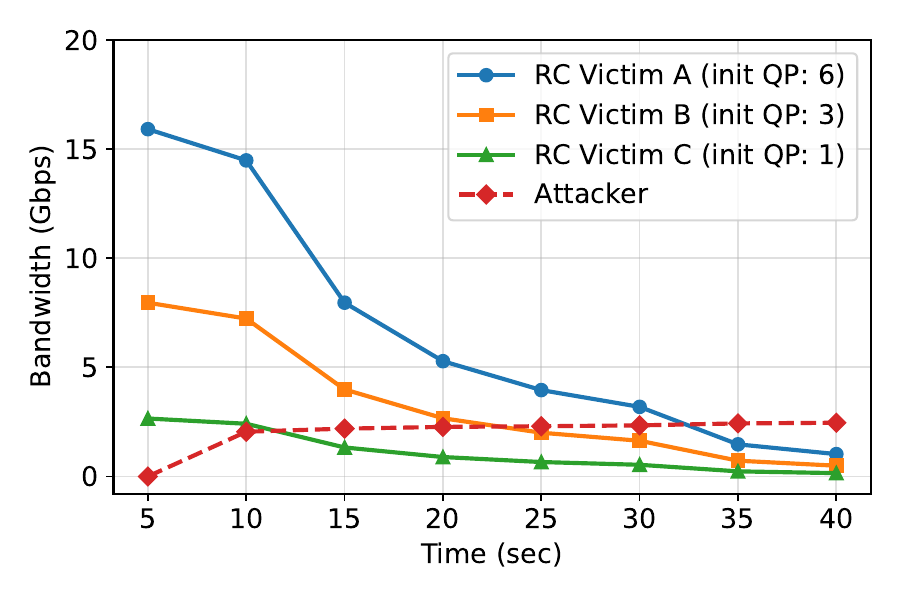}
        \caption{Bandwidth changes in a \emph{multi-victim} environment during a queue flooding attack.}
        \label{fig:queue_flood_result2}
    \end{minipage}
    \vspace{-0.2in}
\end{figure}

\subsection{Effectiveness of Resource Exhaustion Attacks}
\label{sec:Attack Impact analysis}

\noindent\textbf{Impact of Queue Flooding.}
We evaluated the impact of the queue flooding attack by measuring bandwidth degradation experienced by a victim container under both RC and UC modes. Fig.~\ref{fig:queue_flood_result} shows the bandwidth over time for both UC/RC victim and attacker containers during the attack. Under RC mode, the victim's bandwidth dropped from 26.61Gbps to 1.64Gbps, marking a 93.9\% reduction. Under UC mode, a similar degradation trend was observed compared to RC. In contrast, the attacker's bandwidth remained consistently below 5Gbps. These results demonstrate that even lightweight RDMA verbs with minimal data payloads can saturate critical shared RNIC resources, such as TX and RX processing pipelines and WQE cache entries, severely degrading victim performance. This highlights that control intensive verbs, despite their low data volume, are sufficient to monopolize RNIC resources.

To validate the generalizability of this phenomenon in multi-tenant environments, we conducted an additional experiment with three distinct victim containers, each configured with different initial QP counts (6, 3, and 1, respectively). Fig.~\ref{fig:queue_flood_result2} presents the bandwidth changes over time. Initially, Victim A achieved 15.9Gbps, Victim B achieved 7.96Gbps, and Victim C achieved 2.66Gbps. As the flooding continued for 40 seconds, all three victims suffered bandwidth drops of over 90\%. By the end of the 40-second test, the bandwidths fell to 1.02Gbps for Victim A, 0.49Gbps for Victim B, and 0.15~Gbps for Victim C. These results show that queue flooding attacks remain effective even in multi-victim scenarios and demonstrate the need for microarchitectural-level protection mechanisms.

\vspace{0.05in}

\begin{figure}[t]
    \centering
    \begin{minipage}{0.48\linewidth}
        \centering
        \includegraphics[width=\linewidth]{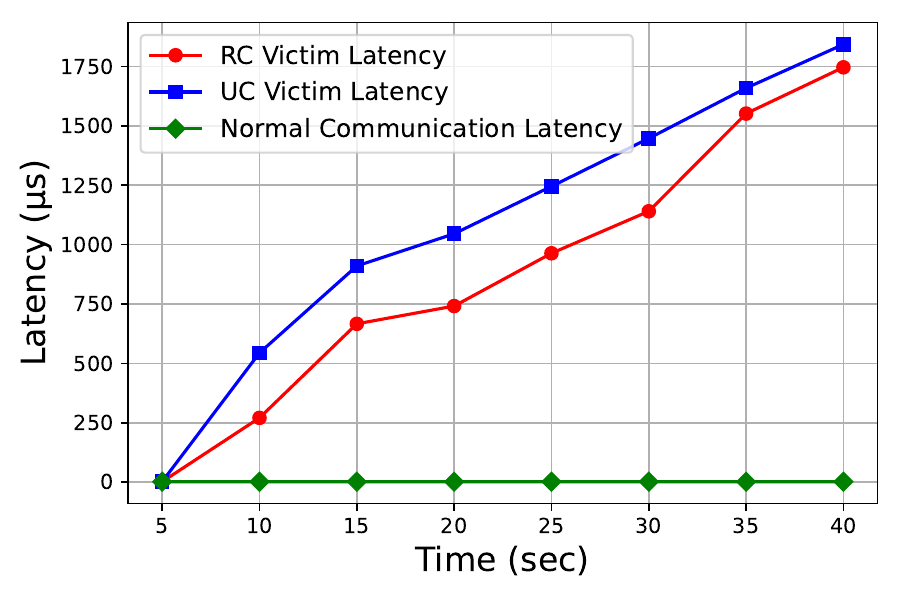}
        \caption{Latency changes of a victim during a \emph{cache depletion attack}.}
        \label{fig:latency_change}
    \end{minipage}
    \hfill
    \begin{minipage}{0.48\linewidth}
        \centering
        \includegraphics[width=\linewidth]{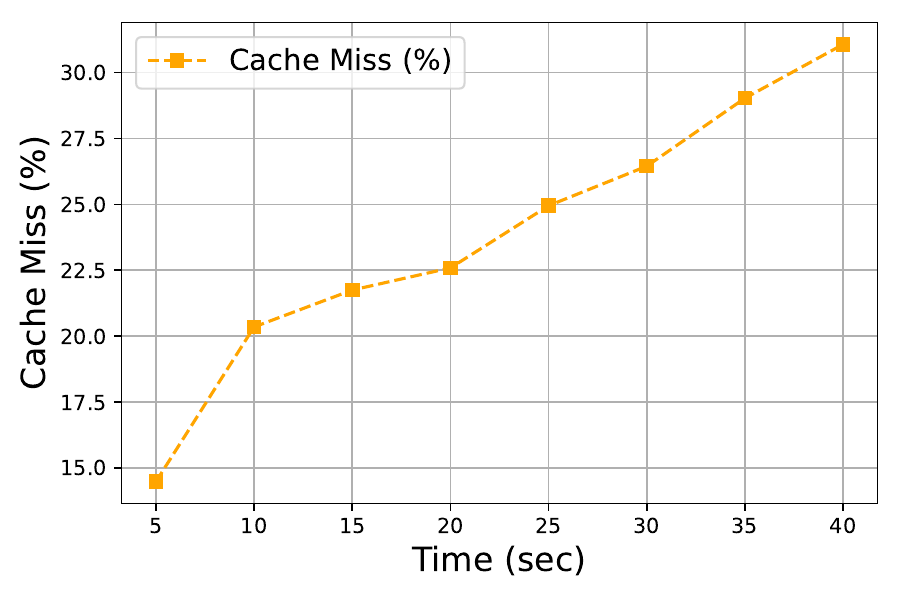}
        \caption{Cache miss rates of a victim during a \emph{cache depletion attack}.}
        \label{fig:cache_miss_change}
    \end{minipage}
    \vspace{-0.2in}
\end{figure}

\noindent\textbf{Impact of Cache Depletion.}
We verified the impact of a cache depletion attack on the victim container by measuring its communication latency and cache miss rates. Under normal conditions, the victim container exhibited an average latency of 1.56~$\mu$s, as shown in Fig.~\ref{fig:latency_change}. During the attack, as the attacker container continuously accessed new memory locations, the victim's latency sharply increased. By the end of the 40-second test, the latency had risen to 1,746.34~$\mu$s, representing approximately a 1,117-fold increase. The cache miss rate also showed a significant rise during the attack, as illustrated in Fig.~\ref{fig:cache_miss_change}. Initially, the cache miss rate was 14.48\%, but it steadily increased throughout the experiment, reaching 31.07\% by the end. These results demonstrate that the cache depletion attack severely degraded the RNIC's cache performance, significantly impacting legitimate RDMA operations.

\subsection{Effectiveness of Pipeline Saturation Attacks}  

\begin{figure}[t]
    \centering
    \begin{minipage}{0.48\linewidth}
        \centering
        \includegraphics[width=\linewidth]{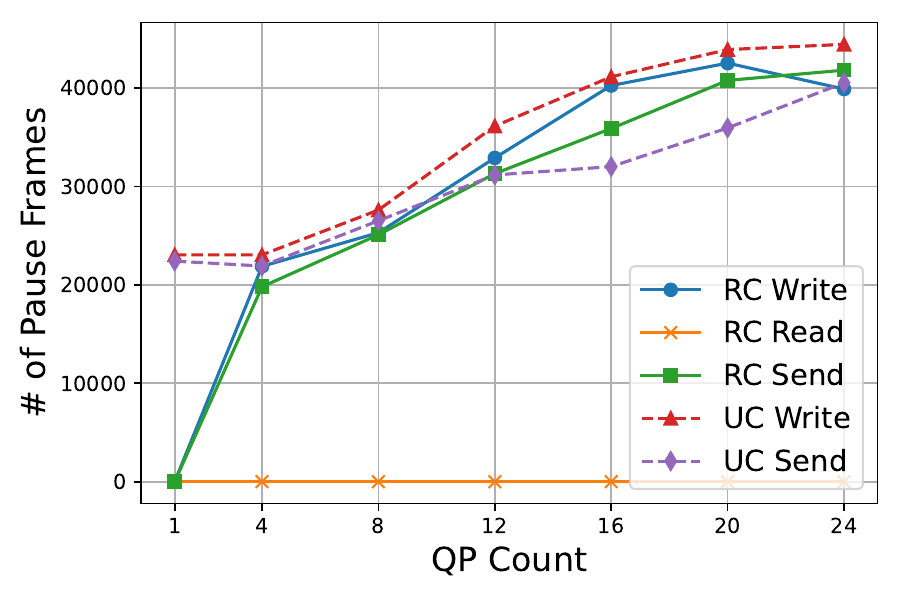}
        \caption{Measured number of PAUSE frames during \emph{verbs flooding attacks}.}
        \label{fig:raf_pause}
    \end{minipage}
    \hfill
    \begin{minipage}{0.48\linewidth}
        \centering
        \includegraphics[width=\linewidth]{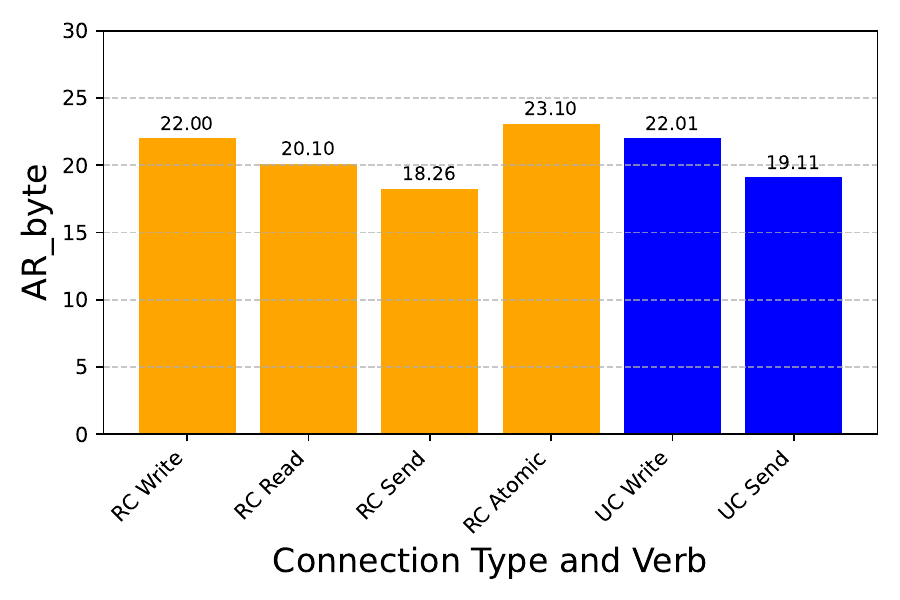}
        \caption{Measured amplification ratio during \emph{verbs amplification attacks}.}
        \label{fig:raf_byte}
    \end{minipage}
    \vspace{-0.2in}
\end{figure}

\noindent\textbf{Impact of Verbs Flooding.} To verify the effectiveness of verbs flooding, we measure the number of PAUSE frames observed during the test. PAUSE frames are flow-control packets that an RNIC sends to its link partner to pause transmission when its RX pipeline or ingress buffers near exhaustion. By counting the number of PAUSE frames, we quantify the RNIC’s link-level congestion during attacks similar to the existing work~\cite{wang2024lordma}. Fig.~\ref{fig:raf_pause} shows the measured number of PAUSE frames when the attacker’s QP count varies from 1 to 24. In RC mode, WRITE verb flooding generates no PAUSE frames at a single QP but escalates rapidly beyond four QPs, reaching a maximum of 42,535 frames at 24 QPs. In contrast, READ verb never triggers PAUSE frames, while SEND verb incurs only modest overhead, rising from 3 frames at one QP to about 40,779 frames at 24 QPs. In UC mode, even a single QP of WRITE or SEND verbs provoke over 22,000 PAUSE frames, climbing to 44,440 and 40,535 frames, respectively as the QP count increases.

\vspace{0.05in}

\noindent\textbf{Impact of Verbs Amplification.} To evaluate the effectiveness of verbs amplification, it is essential to measure how much advantage (i.e., resource consumption) the attacker can gain from sending verb requests. To this end, we define the \emph{amplification ratio (AR)}, which represents the degree to which an RDMA verb $v$ induces load on RNIC resources. Specifically, we define $\mathit{AR}_{\mathrm{byte}}(v)$, which quantifies the number of bytes the RNIC actually receives per byte of a request verb $v$:
\[
  \mathit{AR}_{\mathrm{byte}}(v)
  = \frac{R_v}{L_v},
\]
where $R_v$ denotes the number of bytes received by the RNIC, and $L_v$ indicates the payload size in bytes of the verb $v$. In our experiments, we set $L_v = 8$ bytes and conduct 30-second amplification attacks to measure $\mathit{AR}_{\mathrm{byte}}(v)$.
Here, a high $\mathit{AR}_{\mathrm{byte}}$ value indicates that even small RDMA verbs can impose significant internal load on the RNIC. This overhead results not only from protocol headers and flow control metadata, but also from microarchitectural operations such as queue scheduling, DMA transfers, and cache accesses. As a result, the attacker can consume considerable RNIC resources while keeping bandwidth usage low. This is conceptually similar to amplification attacks like DNS~\cite{ANAGNOSTOPOULOS2013475}, where a small DNS request (e.g., 60 bytes) can trigger responses up to 4,000 bytes, yielding amplification factors exceeding 60$\times$. Likewise, $\mathit{AR}_{\mathrm{byte}}$ reflects how efficiently an attacker can amplify internal RNIC resource usage per byte of issued RDMA verbs, even without increasing overall traffic volume.

Fig.~\ref{fig:raf_byte} illustrates the measured $AR_{byte}$ for each RDMA verb under two different transport modes (i.e., RC and UC). In RC mode, the ATOMIC verb yields the highest amplification ratio (23.1), meaning that the RNIC receives 23.1 bytes for every 8 bytes generated by the attacker, followed by WRITE (22.01), READ (20.1), and SEND (18.26). This result indicates the increasing header and control overhead associated with each verb: ATOMIC incurs both request and response metadata; WRITE adds an extended RDMA header on top of the base transport header; and READ/SEND include only the minimal base header with lightweight control handshakes. The near-parity between RC and UC modes suggests that most of the overhead—such as large header processing, internal metadata handling, and DMA interactions—is imposed by the RNIC microarchitecture itself, regardless of end-to-end reliability. Furthermore, the pronounced gap between one-sided and two-sided verbs corresponds to their protocol complexity: two-sided verbs require additional remote DMA setup metadata, while one-sided verbs rely only on basic network headers and queue handshakes, resulting in lower per-byte amplification.

\section{Discussion and Mitigation}
In this section, we discuss existing solutions and propose potential approaches to detect and mitigate the effects of resource exhaustion attacks targeting RNIC resources in container environments.

\subsection{Root Causes}
The feasibility of RDMA-based resource contention attacks in containerized environments stems from the fact that, under SR-IOV virtualization, all containers share the same set of low-level RNIC microarchitectural resources—such as the Memory Translation Table (MTT), Memory Protection Table (MPT), Connection Cache, WQE cache, and the TX/RX processing pipelines—without any fine-grained partitioning or per-container enforcement. Since SR-IOV assigns only logical VFs to containers, these VFs still contend for on-chip buffers and caches in a first-come, first-served manner. As a result, an attacker that aggressively issues RDMA verbs can evict or throttle legitimate work queue entries and connection state belonging to co-resident victim containers.

Commodity RNICs lack any fine-grained fairness or abuse detection, employing only best-effort or round-robin scheduling across QPs. An attacker can therefore inflate QP counts or flood RDMA verbs to overwhelm the TX/RX pipelines and translation caches driving up cache misses and PCIe back-pressure such as PAUSE frames. While the absence of per-verb rate limiting means sustained verb floods automatically amplify internal load with minimal effort. Because RDMA bypasses the host OS networking stack entirely, container runtimes and OS-level QoS controls cannot observe or throttle these flows. This combination of shared microarchitectural contexts, no scheduling fairness, amplification effects, and kernel bypass creates ideal conditions for devastating resource exhaustion attacks in containerized clouds.

\subsection{Existing Countermeasures}

\noindent\textbf{Hardware-based Solution.} Quality of Service (QoS) mechanism~\cite{NvidiaQosETS} supported by BlueField-3's enables network flow prioritization and resource management by mapping user-defined priorities to traffic classes (TCs). This is achieved through a multi-level process that ensures bandwidth guarantees, restrictions, or prioritized access to network resources. For example, the bandwidth speed for ports allocated to tenants can be limited using commands such as:  \texttt{echo 1000 > /sys/class/net/p0/\{BlueField\}/\{pf,vf\}/}\texttt{\{min,max\_tx\_rate\}}. This configures the maximum and minimum transmission speeds for PFs and VFs, ensuring fair resource allocation among tenants. Additionally, tools such as \texttt{mlxdevm} and \texttt{evlink} enable per-port or per-group bandwidth settings. Furthermore, the Enhanced Transmission Selection (ETS)~\cite{NvidiaQosETS} mechanism enables dynamic bandwidth sharing, maintaining overall system performance by allocating unused bandwidth effectively. 

On the other hand, \texttt{VirtIO-net}~\cite{VirtioNet} enables static QP provisioning per VF, offering hardware-level configuration flexibility, it was originally designed for QEMU-based virtual machine environments. Consequently, it operates at the PCIe device level and lacks native support for dynamic attach/detach and fine-grained resource isolation, making it ill-suited for modern container environments such as Docker. This architectural constraint limits its usability in multi-tenant containerized systems, where per-container flexibility and strict performance isolation are required.

\begin{figure}[t]
    \centering
    \begin{minipage}[t]{0.48\linewidth}
        \centering
        \includegraphics[width=\linewidth]{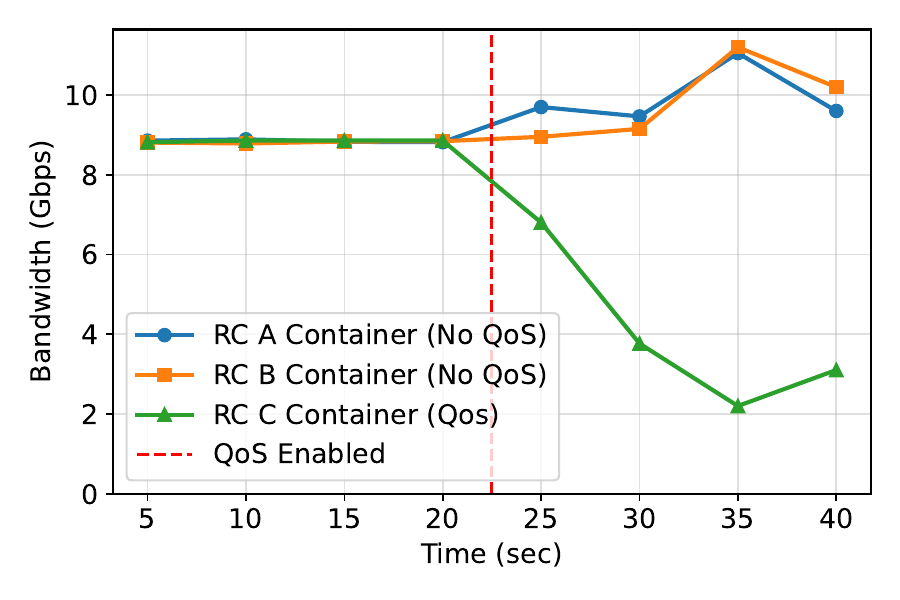}
        \caption{Impact of QoS enforcement on RDMA bandwidth distribution among containers.}
        \label{fig:z}
    \end{minipage}
    \hfill
    \begin{minipage}[t]{0.48\linewidth}
        \centering
        \includegraphics[width=\linewidth]{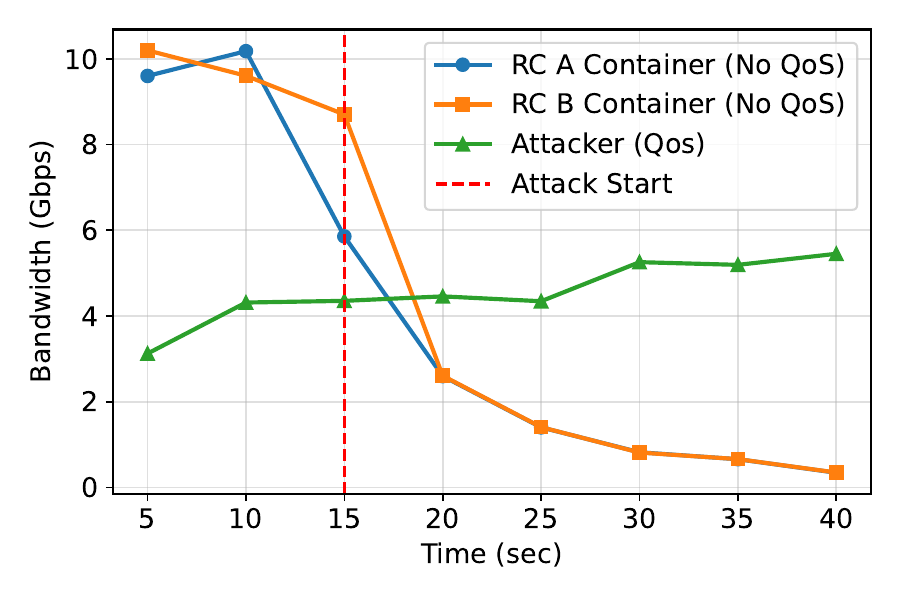}
        \caption{(In)effectiveness of the QoS mechanism in limiting the attacker container during a \emph{queue flooding attack}.}
        \label{fig:z2}
    \end{minipage}
    \vspace{-0.2in}
\end{figure}

While these all mechanisms are effective for managing network resources at the traffic level, they face limitations in addressing performance isolation issues specific to RNICs. For example, NVIDIA's RoCE protocol supports only four predefined ToS (Type of Service) values, limiting the flexibility of priority configurations in complex multi-tenant environments. Consequently, QoS mechanisms often struggle to adapt to dynamically changing network traffic patterns.  Moreover, since QoS mechanisms are primarily designed to regulate traffic prioritization, they cannot directly mitigate resource exhaustion attacks targeting RNIC resources such as the TX/RX processing unit or internal caches.

To validate this limitation, we conducted an experiment using three containers, each subject to QoS mechanisms. We configured one container with an Enhanced Transmission Selection (ETS) weight of 10\% and a 4Gbps rate limit, then measured the bandwidth distribution before and after enforcing the policy. As shown in Fig.~\ref{fig:z}, once the QoS policy was applied at 23 seconds, it effectively capped Container C’s bandwidth, enabling Containers A and B to opportunistically utilize the freed RNIC capacity. Fig.~\ref{fig:z2} extends this analysis by applying the same QoS configuration to the attacker container during a queue flooding attack. Despite throttling the attacker at 15 seconds, Containers A and B—without proper isolation—still suffered a 73.4\% performance degradation compared to the baseline. These results demonstrate that although ETS can redistribute bandwidth, it does not guarantee minimum service levels in the absence of microarchitectural or verb-level isolation.

\vspace{0.05in}

\noindent\textbf{Software-based Solution.} Several software-based solutions have been proposed to prevent resource exhaustion attacks. Freeflow~\cite{kim2019freeflow} is a representative system designed to ensure fair resource allocation and predictable performance in RDMA-enabled container environments. Freeflow achieves its goals by creating virtual NICs and virtualized spaces, which allow transparent management of resource movement. By virtualizing the resources and managing them dynamically, Freeflow ensures that containers can access resources fairly without interference. While Freeflow provides a transparent and structured methodology for resource allocation, its resource management approach may not fully address bottlenecks at the microarchitecture level, such as contention in compute or I/O units, which can significantly impact performance in high-demand scenarios. In addition, prior work has shown that such software-based virtualization introduces overheads in the RDMA data path, which counteract RDMA’s performance benefits~\cite{li2024dockrdma}.

\subsection{\ourtool{}: A Threshold Based Resource Management System}
\begin{figure}[t]
    \centering
    \includegraphics[width=1\linewidth]{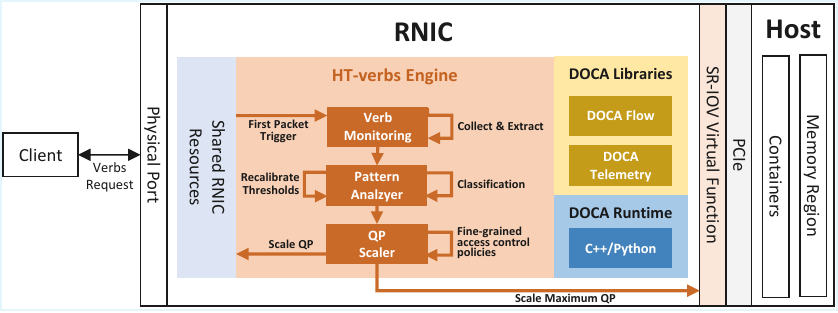}
    \caption{The conceptual architecture of \ourtool{}.}
    \label{fig:Fig5}
    \vspace{-0.2in}
\end{figure}

To overcome the limitations of existing resource isolation mechanisms and mitigate resource exhaustion attacks in RDMA-enabled container environments, we propose \ourtool{}, a threshold-based RNIC resource management system. Fig.~\ref{fig:Fig5} illustrates its conceptual architecture. \ourtool{} operates on the RNIC to dynamically monitor and classify per-QP resource usage in real-time, enabling the system to detect malicious QPs. Upon detection, \ourtool{} can selectively throttle or block malicious QPs at the SmartNIC level, while Adaptively scaling legitimate QPs to ensure fair and predictable performance across legitimate containers

\vspace{0.05in}

\noindent\textbf{Resource Monitoring.}
\ourtool{} periodically collects and aggregates the usage patterns of RDMA Verbs (e.g., READ, WRITE, SEND, RECV, ATOMIC) invoked on each QP. Key metrics, such as QP/CQ creation frequencies, TX/RX pipeline pressure, and cache hit rates, are continuously monitored within the RNIC's shared microarchitectural layer. These metrics are collected and tracked through the \textit{Verb Monitoring} component of \ourtool{}, and the engine further consists of two additional modules—\textit{Pattern Analyzer}, which detects abnormal usage trends per container and analyzes QP usage patterns in accordance with dynamic temperature levels; and \textit{QP Scaler}, which dynamically adjusts resource allocation based on pattern analyzer outcomes. These three components operate in tandem to identify potential resource abuse and ensure fair resource sharing across multi-tenant environments.

\vspace{0.05in}

\noindent\textbf{Resource Classification.}
Based on the analysis of usage patterns, \ourtool{} categorizes RNIC resources into three levels: \textit{Hot}, \textit{Warm}, and \textit{Cold}, with each classification representing a different level of resource demand:
\begin{itemize}
    \item \textbf{\emph{Hot}:} These are heavily utilized resources experiencing high contention. Access by lower-priority or malicious containers is restricted to prevent overloading.
    \item \textbf{\emph{Warm}:} These resources have moderate demand, maintaining standard isolation levels. Both legitimate and lower-priority containers can access these resources with minimal restrictions.
    \item \textbf{\emph{Cold}:} These are underutilized resources. Higher-priority containers are granted enhanced performance, exceeding typical isolation guarantees.
\end{itemize}
This classification is performed by the \textit{Pattern Analyzer} and provided to the \textit{QP Scaler} as input for enforcing appropriate resource controls.

\vspace{0.05in}

\noindent\textbf{Adaptive Threshold Adjustment.}
The threshold criteria for classifying resources into \textit{Hot}, \textit{Warm}, and \textit{Cold} are not static. Instead, they are continuously recalibrated by the \textit{Pattern Analyzer} based on system-wide metrics collected through the \textit{Verb Monitoring} module, such as overall RNIC utilization, traffic fluctuation, and per-QP statistics. These adaptive thresholds are used by the \textit{QP Scaler} to enforce dynamic resource control, allowing \ourtool{} to remain responsive to sudden demand spikes or evolving attack behavior while preserving performance stability.

\vspace{0.05in}

\noindent\textbf{Malicious Container Detection.}
\ourtool{} does not rely on fixed rules or static signatures to identify malicious containers. Instead, it leverages the \textit{Verb Monitoring} module to track per-QP RDMA usage patterns in real time, which are then evaluated by the \textit{Pattern Analyzer} against system-wide activity distributions. A container is considered anomalous if it consistently deviates from normal behavior—for example, through excessive QP/CQ creation or persistently high TX/RX pipeline pressure without a proportional increase in throughput. These classification results are forwarded to the \textit{QP Scaler}, which selectively restricts or deprioritizes the offending container's resource access. In low-contention scenarios or when only a single container is active, threshold constraints are relaxed to avoid penalizing benign high-performance workloads.

\vspace{0.05in}

\noindent\textbf{Implementation Strategy.} \ourtool{} operates as a middleware layer between the RNIC hardware and containerized applications. It can be implemented atop NVIDIA’s BlueField-3 architecture using the DOCA SDK~\cite{doca_sdk}, which provides low-level telemetry access and control APIs for RDMA-aware SmartNICs. Specifically, the \textit{Verb Monitoring} module leverages DOCA Telemetry~\cite{DocaTel} and \texttt{mlx5} hardware counters to periodically collect per-VF resource metrics—such as QP/CQ creation rate, TX/RX pipeline occupancy, and cache activity—at fine-grained intervals (e.g., every 100 ms). Container association is achieved using the \texttt{pipework} utility~\cite{petazzoni2014pipework}, which binds VFs to container network namespaces, or alternatively via DOCA’s container-awareness hooks.  These raw metrics are streamed to the \textit{Pattern Analyzer}, which uses statistical analysis to compute both absolute and differential indicators (e.g., ΔPAUSE frames/sec, cache miss rate). Thresholds are dynamically adjusted based on recent statistical distributions (e.g., 90th percentile = ``Hot'', 10th percentile = ``Cold''). 

These modules can be implemented using standard Python/C++ bindings atop the DOCA Telemetry library for real-time analysis within the DPU OS. Based on classification results, the \textit{QP Scaler} enforces per-container policies using DOCA Flow APIs~\cite{DocaFlow}. Additionally, it enforces fine-grained access control policies, such as throttling (via WQE pacing), priority adjustment, or QP reallocation. Policy enforcement is performed inline on the DPU to minimize control-plane delay. Throughout this process, \ourtool{} continuously monitors feedback signals such as throughput oscillation and WQE latency. Thresholds and control policies are recalibrated in real time to ensure stability under multi-tenant contention.
\section{Related Work}

\noindent\textbf{Performance Isolation in RDMA.}
Performance isolation in RDMA-enabled environments has been extensively explored, primarily focusing on virtual machines and bare-metal systems. Kong et al.~\cite{kong2023understanding} conducted a comprehensive analysis showing how RDMA microarchitectural resources—such as RNIC caches, processing units, and PCIe bandwidth—become critical bottlenecks for performance isolation in multi-tenant clouds. They demonstrated that existing isolation techniques fail to mitigate contention at this level and introduced the Husky test suite to systematically evaluate such issues. Similarly, PeRF~\cite{298513} examined QP-level unfairness in RNICs, revealing how round-robin scheduling can cause throughput collapse and latency spikes for low-QP tenants. Unlike these studies, our work focuses on performance isolation challenges specific to RDMA-enabled container environments.

\vspace{0.05in}

\noindent\textbf{Solutions for Performance Isolation.}
To address the performance isolation, many systems have been proposed. Harmonic~\cite{lou2024harmonic} introduced hardware-assisted mechanisms to ensure performance isolation in public clouds. OSMOSIS~\cite{khalilov2024osmosis} introduced a hardware-software co-design that isolates DMA queues, execution pipelines, and internal scheduling domains within SmartNICs. FairNIC~\cite{grant2020smartnic} proposed static partitioning of SmartNIC cores, buffers and caches and applies a distributed token based rate-limiter for on-board accelerators, thereby eliminating cross-tenant contention while sustaining the Cavium LiquidIO SmartNIC's full 25 Gbps line-rate throughput. multi-tenants. In this paper, we propose \ourtool{}, which collects key metrics derived from the requested RDMA verbs and achieve performance isolation per container through dynamic adaptive throttling, without requiring hardware or firmware modifications of an RNIC.

\vspace{0.05in}

\noindent\textbf{RDMA in Container Environments.}
Recently, there has been growing interest in applying RDMA to container environments~\cite{kube_sriov}. Freeflow~\cite{kim2019freeflow} introduced a structured resource allocation approach for SmartNIC-enabled systems by creating virtual NICs and virtualized address spaces. DockRDMA~\cite{li2024dockrdma} proposed a hybrid RDMA virtualization framework that dynamically switches between VF-based direct access and a proxy-based fallback. Liu et al.~\cite{305973} addressed RDMA scalability limitations in containerized networks by dynamically steering traffic between one-sided RDMA and connectionless or proxy-based paths. TSoR~\cite{sun2023tsor} presented a turnkey RDMA network solution for containerized applications, transparently multiplexing POSIX-socket traffic over pre-established RDMA channels and integrating with Kubernetes via a CNI plugin.

\vspace{0.05in}

\noindent\textbf{RDMA Security.}
Several studies have investigated RDMA security and uncovered a range of vulnerabilities. Bedrock~\cite{277184} employed P4-programmable switches to protect RDMA traffic against spoofing, access violations, and side-channel attacks. LoRDMA~\cite{wang2024lordma} identified a class of low-rate DoS attacks that exploit the interaction between priority flow control and DCQCN (Data Center Quantized Congestion Notification), where short, coordinated bursts from multiple senders can induce global congestion and cause DCQCN to throttle benign flows. ReDMArk~\cite{rothenberger2021redmark} analyzed RDMA’s built-in security mechanisms, demonstrating how weak authentication allows access-key inference, memory rekeying, and connection hijacking. These attacks take advantage of RDMA’s insufficient isolation and lack of robust authentication to enable unauthorized memory access across tenants. In contrast, our work assumes an adversary within RDMA-enabled container environments and targets the RNIC’s microarchitectural resources to induce performance interference in other victim containers.

\section{Conclusion and Future Work}

In this work, we experimentally analyzed performance isolation problems in RDMA-enabled container environments and proposed two types of resource exhaustion attacks that target RNIC resources: (i) state saturation attacks and (ii) pipeline saturation attacks. Our experiments reveal that state saturation attacks cause up to 93.9\% bandwidth loss, a 1,117× latency increase, and a 115\% rise in cache misses, while pipeline saturation attacks induce severe link-level congestion and amplification, where small verb requests consume disproportionately large RNIC resources. To counter these threats, we propose \ourtool{}, a threshold-driven framework that, using real-time RDMA verb telemetry, classifies RNIC resources into Hot, Warm, and Cold tiers, dynamically throttling abusive workloads without requiring hardware modifications. As future work, we plan to implement a prototype of \ourtool{} and empirically validate its practicality and effectiveness under real-world container workloads.

\begin{credits}
\subsubsection{\ackname} This work was supported by the National Research Foundation of Korea (NRF) grant funded by the Korea government (MSIT) (No. RS-2024-00457937, Design and implementation of security layers for secure WebAssembly-based serverless environments).
\end{credits}

%
%
%
\bibliographystyle{splncs04}
\bibliography{references}

\end{document}